\documentclass[preprint,aps,amsmath,amssymb,nofootinbib,showpacs,
showkeys,tightenlines,secnumarabic]{revtex4}
\usepackage{graphicx}
\newcommand{\ba}{\begin{eqnarray}}
\newcommand{\ea}{\end{eqnarray}}
\newcommand{\nn}{\nonumber}

\begin{document}

\title{First-passage method for the study of the efficiency  
of a two-channel reaction on a lattice}
\author{E. Abad} 
\email{eabad@ulb.ac.be}
\affiliation{Centre for Nonlinear Phenomena and Complex Systems, 
Universit\'e Libre de Bruxelles CP 231, 1050 Bruxelles, Belgium.}

\begin{abstract}
We study the efficiency of a two-channel reaction between two walkers
on a finite $1D$ periodic lattice. The walkers perform a combination of
synchronous and asynchronous jumps on the lattice and react instantaneously
when they meet at the same site (1st channel) or upon position exchange (2nd
channel). 
We develop a method based on a conditional first-passage problem to 
obtain exact results for the mean number of time steps 
needed for the reaction to take place as well as for higher order
moments. Previous results 
obtained in the framework of a difference equation approach are 
fully confirmed, including 
the existence of a parity effect. For even
lattices the maximum efficiency corresponds to a mixture of 
synchronous events and a small amount of asynchronous events,
while for odd lattices the 
reaction time is minimized by a purely synchronous process. 
We provide an intuitive explanation for this behavior. 
In addition, we give explicit expressions for the variance of the 
reaction time. The latter displays a similar even-odd
behavior, suggesting that 
the parity effect extends to higher order moments. 
\end{abstract}

\maketitle

\section{Introduction}

In experimental systems there is always a finite 
discretization of the time window used for measurements.
Thus, the degree of synchronicity for the dynamics of the different system 
constituents may vary when the size of the time window is
changed, either intentionally or accidentally as a result 
of random fluctuations in the clock mechanism. But even for 
a fixed time resolution there may be 
intrinsic features of the experimental system that lead to the 
coexistence of synchronous and asynchronous
events in the course of the evolution. Consider e.g. 
diffusion of an ensemble of identical particles in a 
disordered or a 
randomly fluctuating medium \cite{bouch,benhav}. The diffusivity 
of a particle will 
then be different depending on its location, and this will lead to 
the coexistence of different characteristic time scales in the system, 
resulting in partial desynchronization for the dynamics of its constituents. 


Recently, the author and coworkers used a simplified lattice version of 
this problem to study the role of synchronicity on the 
efficiency of a diffusion-controlled two-channel reaction between 
two particles 
\cite{abad}. Here, the parameter chosen for the characterization 
of the reaction efficiency is the mean number of time steps 
to reaction. The effect of synchronicity was investigated by studying
how this quantity is affected by the interplay 
between the transport dynamics and the geometric characteristics of the 
lattice, i.e. size, dimensionality and boundary conditions. Finite size 
effects were found to play an important role, thus emphasizing the 
relevance of geometric constraints for systems with a small number 
of constituents.
On the other hand, a variety
of interesting problems in statistical physics may be recast in terms 
of a random walk in a lattice of a small size. An example thereof 
is a family of ruin problems where
the capital of each gambler is typically a small number \cite{feller,abad}.  

In ref. \cite{abad} the analytic results for the $1D$ case were based 
on a difference equation 
approach, whereas numerical simulations were used in higher dimensions. 
However, it is also instructive to compute the mean reaction time 
for this problem by 
other methods, e.g. approaches 
based on the theory of finite Markov processes \cite{kem,bentz,these}
and on generating function techniques \cite{weiss}. 
The first method has the advantage 
of being applicable 
in higher dimensions and to systems lacking translational invariance 
(e.g. with reflecting boundaries), while 
the second emphasizes the correspondence between
diffusion-controlled reactions and first-passage problems \cite{redn}.  
In this paper, we shall apply the latter to determine the
mean number of time steps necessary for the reaction to 
take place as well as the variance of this quantity. 
An important advantage of this approach is that the variance 
and higher order moments can be computed
by a straightforward differentiation of the relevant generating function,
as opposed to the method of difference equations, where moments 
are coupled to each other via a hierarchy of equations \cite{abad}. 

The work plan is as follows: in Section \ref{themod} 
we introduce the model and its reduced representation in
a comoving reference frame.  
Section \ref{genfunap} introduces the generating 
function approach to compute the reaction time and higher order moments. 
For completeness, we first present the standard formalism to deal 
with some simple cases where only a single reaction channel
is active. We then treat the general case, which
is solved in two stages. First, we show that our model
is equivalent to a model with a relaxed collision rule and then
we solve for the latter. Section \ref{discuss} contains the main 
results for the reaction time and its variance.
Finally, Section \ref{concl} summarizes the main conclusions. 

\section{The model}
\label{themod}

The system to be studied consists of two coreactants $A$ and $B$
performing symmetric nearest neighbor random jumps on a $N$-site 
periodic lattice at discrete time
steps. The reactants, also termed walkers in what 
follows, are assumed to react with 
each other whenever they meet at the same lattice site or attempt to
exchange positions. Each of such ``collisions'' results in an
instantaneous irreversible reaction. 
Regardless of the particular outcome of the reaction \footnote{The case
$A+B\rightarrow inert.$ is of great historical interest,
see e.g. refs. \cite{Ovch,tous,soko,lind,kang,kang2}}, the mean reaction time 
can be identified with the mean number of time steps elapsed 
until the collision
takes place; the larger this quantity, the less efficient the 
reaction will be. Each time step 
will be considered to be an elapsed time unit regardless of the 
lattice size. 
In the course of the dynamics, the following joint events 
may occur:  

\begin{enumerate}

\item with probability $p$ both walkers hop simultaneously 
to randomly chosen nearest neighbor sites (synchronous event).

\item with probability $1-p$ one of the walkers (no matter whether $A$ or $B$) 
hops to a nearest neighbor site while the other one remains immobile 
(asynchronous event). 


\end{enumerate}

Thus, the characteristic parameter $p$ interpolates 
between the asynchronous case ($p=0$) 
and the case of two simultaneously moving walkers ($p=1$). 
Occasionally, we shall refer to the limiting cases $p=0$ and $p=1$
as ``the purely asynchronous case'' and ``the purely synchronous case'' 
respectively. The purely asynchronous case was studied by Montroll 
\cite{mont0,mont1}
and Montroll and Weiss \cite{monwei} in one, two and three dimensions.
For the particular case of a periodic $1D$ lattice they obtained: 

\begin{equation}
\label{purasc}
\langle n\rangle =\frac{N (N+1)}{6},
\end{equation}
where $\langle n \rangle$ is the initial-condition-averaged
mean reaction time. Note that in this case the only
active reaction channel is same site occupation (SSO), 
since reaction by position exchange of the walkers, i.e. by
nearest neighbor crossing (NNC) is not possible.
 
However, in the purely synchronous case, both reaction channels SSO and NNC 
will be active if the total number of sites $N$ is odd, otherwise only
one channel will be available, depending on the initial location of
the walkers. A consequence is that the analytical expression
for $\langle n \rangle$ in terms of the
size of the $1D$ lattice depends on the parity of $N$ \cite{bentz,nickoz1}. 
This even-odd effect translates mathematically as follows \cite{abad}: 

\begin{equation}
\label{evenoddenct}
\langle n\rangle=
\left\{ \begin{array}{cc}
N(N+1)(N+2)/(12\,(N-1)) & \qquad \mbox{for $N$ even, } \\
(N+1)(N+3)/12 & \qquad \mbox{for $N$ odd. } 
\end{array} \right. 
\end{equation}
In contrast to the above cases, 
as soon as $0<p<1$, reaction by NNC becomes possible 
{\it regardless} of the value of $N$ and the initial 
two-walker configuration.  

\begin{figure}[ht]
\includegraphics[width=14cm,height=6cm]{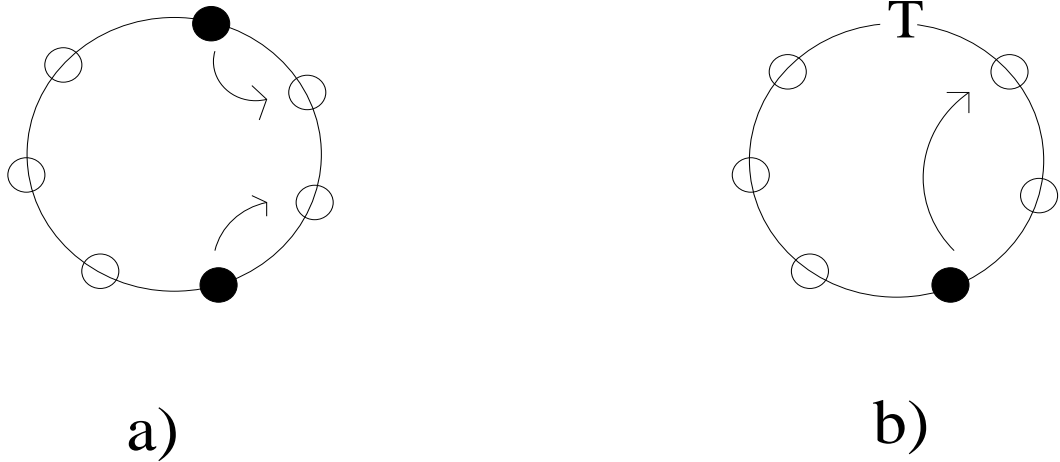}
\vspace{.5cm}
\caption{\it \label{felfig1} a) Two-walker system on a seven-site 
periodic lattice. Both walkers are represented by black circles. 
For convenience, the walker labels $A$ and $B$ have been left out
(see text for explanation). b) Equivalent one-walker plus trap system.
The trapping site is denoted by ``T''.  
The arrows in fig. \ref{felfig1}a indicate that both walkers perform 
a synchronous step. In fig. \ref{felfig1}b this corresponds to a 
two-site jump of the walker.}
\end{figure}

Due to the translational invariance of the lattice,
only the relative motion of 
both walkers is relevant for the computation of the collision time.
This has two important consequences. First, it tell us that the physical 
distinguishability of the walkers is irrelevant for the 
solution of the problem. 
To emphasize this, the walker labels $A$ and $B$ have been left out
in the scheme displayed in fig. \ref{felfig1}a. The second consequence 
is that it is convenient to choose the reference frame in such a way 
that one of the walkers (say $B$) is at rest. In this comoving frame, 
walker $A$ will either hop to a nearest neighbor (with probability $1-p$),
hop to a {\it next to nearest neighbor} site (with probability $p/2$)
or remain immobile (again with probability $p/2$) as a result of 
the rules 1 and 2 prescribed above. 
In this one-walker representation, 
walker $B$ plays the role of a stationary trap $T$, 
as indicated in fig. \ref{felfig1}b; any time walker $A$ reaches {\it or} 
attempts to overcome the site at 
which walker $B$ is placed, the instantaneous reaction is triggered
and the dynamics is stopped. In what follows, we shall therefore
refer to walker $B$ as ``the trap'' and to walker $A$ as
``the walker'' when we work in the comoving frame.

For convenience, let us place the origin (site $0$) of the 
comoving frame at the initial position 
of walker $A$ and then number the remaining sites, say clockwise, 
from $1$ to $N-1$. Denoting by $j_T$ the coordinate of the site at which 
the trap is located, the distance in lattice spacings between
both walkers will be $d=\min(j_T,N-j_T)$.   

\section{Generating function approach}
\label{genfunap}

Our next goal will be to derive an expression for the mean number of 
time steps ${\langle n \rangle}_{j_T}$ necessary for the reaction
to take place for a given value of the coordinate $j_T$
characterizing the initial condition.
As a starting point, we take the equations that would govern
the sojourn probabilities of an unrestricted walk {\it if} there were
no interaction between the walker and the trap. These equations read  

\begin{equation}
\label{pndyn}
{\cal P}_{n+1}(j)=\frac{p}{4}\,\left[ {\cal P}_n(j-2)+
2{\cal P}_n(j)+{\cal P}_n(j+2) \right]
+\frac{1-p}{2}\,\left[ {\cal P}_n(j-1)+{\cal P}_n(j+1)\right]
\end{equation}
where ${\cal P}_n(j)$ is the probability
to find the walker at a given site $j$ after $n$ time steps
($j$ may take integer values from $0$ and $N-1$ 
and the site addition and subtraction is performed modulo $N$). 
The first term in the r.h.s. of eqs. (\ref{pndyn}) 
is the contribution due to the 
synchronous events, by which the walker either remains at rest 
or it moves two lattice sites clockwise or anticlockwise. The second term 
describes jumps by one lattice site yielded by the asynchronous events.  
In accordance with our definition for the origin, eqs. 
(\ref{pndyn}) must be solved using the deterministic initial condition 
${\bf P}_0=({\cal P}_0(0),{\cal P}_0(1),\ldots,{\cal P}_0(N-1))^T= 
(1,0,\ldots,0)^T$. 






\subsection{A simple case}

The next step is to incorporate the walker-trap interaction to 
the above formalism. Let us first consider the situation where 
reaction by NNC is precluded. This holds if and only if $N$ 
and the walker-trap separation $d$ are even integers.  In this
case site $j_T$ can be 
viewed as a reactive site, also termed ``$r$-site''
in what follows. Clearly, the mean reaction time will be given 
by the mean first-passage time ${\langle n \rangle}_{j_T}$ of the walker 
at site $j_T$:

\begin{equation}
\label{fpast}
{\langle n \rangle}_{j_T}=\sum_{n=1}^\infty n\,{\cal F}_n(j_T),
\end{equation}
where ${\cal F}_n(j)$ is the probability of visiting a given site $j$ 
for the {\it first} time after $n$ time steps. Eq. (\ref{fpast}) can be 
expressed as follows

\begin{equation}
\label{enctjB}
{\langle n \rangle}_{j_T}=\left.\frac{\partial}{\partial z}
{\cal F}(j_T,z)\right|_{z=1},
\end{equation}
where ${\cal F}(j,z)\equiv\sum_{n=0}^\infty {\cal F}_n(j)\,z^n$
is the generating function of the first-passage probabilities
${\cal F}_n(j)$ and the limit $z\rightarrow 1$ is taken from
below. On the other hand, this function can be
directly related to the generating function  
${\cal P}(j,z)\equiv\sum_{n=0}^\infty {\cal P}_n(j)\,z^n$. 
To do so, one uses the fact that the sets of 
probabilities ${\cal P}_n(j)$ and ${\cal F}_n(j)$ 
are linked to each other via the equation \cite{weiss}

\begin{equation}
\label{convsum}
{\cal P}_n(j)=\sum_{k=1}^n {\cal F}_k(j) {\cal P}_{n-k}(0), \qquad j\neq 0.
\end{equation}
The discrete convolution in the r.h.s. of eq. (\ref{convsum}) 
is equivalent to the product ${\cal P}(j,z)\,{\cal F}(j,z)$ 
in the reciprocal generating function space. Thus, we have

\begin{equation}
\label{gennotor}
{\cal F}(j,z)=\sum_{n=0}^\infty {\cal F}_n(j)\,z^n 
=\frac{{\cal P}(j,z)}{{\cal P}(0,z)},\quad j\neq 0.  
\end{equation}
The collision time ${\langle n \rangle}_{j_T}$ then follows from eqs. 
(\ref{enctjB}) and (\ref{gennotor}): 

\begin{equation}
\label{defenct}
{\langle n \rangle}_{j_T}=\left.\frac{\partial}{\partial z}\,
\frac{{\cal P}(j_T,z)}{{\cal P}(0,z)} 
\right|_{z=1}.
\end{equation}

\subsection{General case}

Let us now extend these results to the general case
where $N$, $d$ and the synchronicity parameter $p$ take 
arbitrary values. The strategy to 
tackle the problem will be as follows: we shall not deal with 
NNC events directly, but rather introduce
a model with a relaxed definition of collision, show 
its equivalence to the original one
by virtue of the topological restrictions imposed by the $1D$ lattice
and then compute the reaction time for the relaxed model. 
For brevity, let us respectively refer to the original and the 
relaxed models as ``model I'' and ``model II'' in what follows. 

\begin{figure}[ht]
\includegraphics[width=4cm,height=4cm]{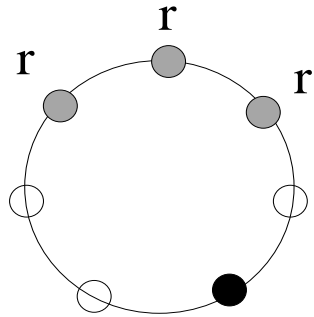}
\caption{\it \label{modmorep} Representation of model II 
in the comoving frame.}
\end{figure}

\begin{figure}[ht]
\includegraphics[width=14cm,height=4cm]{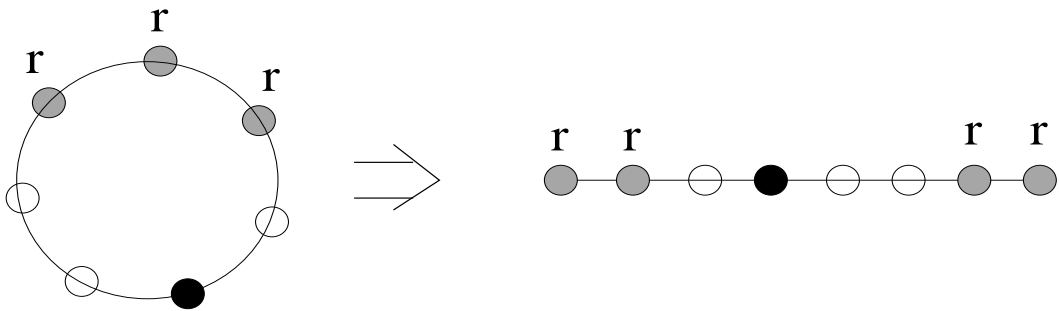}
\caption{\it \label{modunf} Lattice transformation for 
model II in the comoving frame. A periodic lattice 
with three reactive sites is equivalent to a non-periodic one
with four $r$-sites}
\end{figure}

In model II, one assumes that the walkers react instantaneously
by SSO {\it or} when they jump to {\it nearest neighbor sites}, i.e.
by nearest neighbor occupation (NNO). The representation of the system
in the comoving frame will consist of a walker in a lattice 
with three $r$-sites, as shown in fig. \ref{modmorep} for a 
7-site system. According to the site numbering
introduced in Sec. \ref{themod},
these three $r$-sites will have the coordinates
$j_T$ (where the immobile reactant is located), $j_T-1$ and $j_T+1$.
By definition, the walk will automatically terminate when the walker
lands on any of the three $r$-sites. For the purpose of computing 
$\langle n \rangle_{j_T}$, it is easy to 
realize that the system can be unfolded into an equivalent
system with a non-periodic lattice 
by introducing an additional fictitious $r$-site, 
as shown in fig. \ref{modunf} for $N=7$. 



\begin{figure}[ht]
\includegraphics[width=8cm,height=7cm]{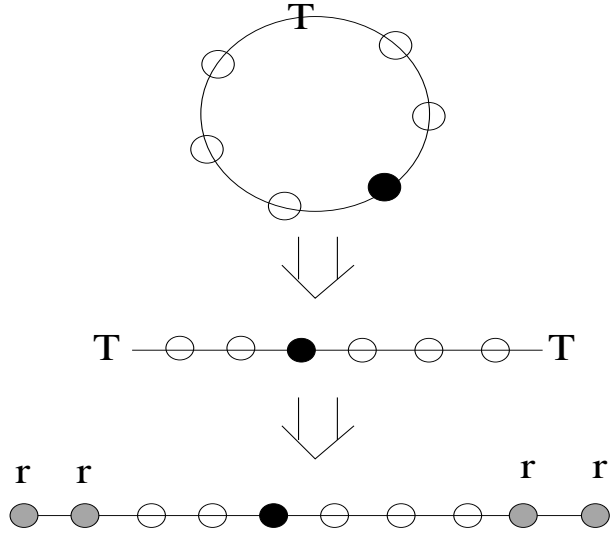}
\caption{\it \label{felfig2} Lattice transformation for the 
one-walker representation of model I displayed in fig. \ref{felfig1}b.}
\end{figure}

On the other hand, let us again 
consider the walker-trap representation of
the model I depicted in fig. \ref{felfig1}b for $N=7$.
Along the same lines as above, the system can be unfolded into
an equivalent transformed lattice with two trapping sites $T$ and 
$N-1=6$ non-trapping sites (cf. fig. \ref{felfig2}). 
Each site $T$ can then be replaced with two fictitious 
$r$-sites, as shown in fig. \ref{felfig2} \footnote{
If the dynamics is purely 
asynchronous ($p=0$), only one $r$-site at each end will be needed, 
since jumps by two sites are not possible in this case.}.  
Thus, we realize that
model I embedded in a periodic lattice with $N$ sites is equivalent to 
model II in a periodic lattice with $N+2$ sites. 



Next, let us compute the reaction time $\langle n\rangle_{j_T}$ for model II.
The situation is now more complex than the one described in the 
previous subsection, since we do not know {\it a priori} at which of the 
three $r$-sites 
$j_T-1$, $j_T$ or $j_T+1$ the reaction will occur. Yet, 
it can still be formulated as a (conditional) first-passage problem. 
The key quantity in this case is ${\cal F}_n(s\,|\, s_1,n_1; s_2, n_2)$, 
i.e. the probability that a random walker arrives at site $s$ for the 
{\it first} time at the 
$n$-th time step after having visited the sites $s_1$ and 
$s_2$ exactly $n_1$ and $n_2$ times respectively. For a given initial 
condition with a fixed value of $j_T$, the mean collision time is now 

\ba
{\langle n \rangle}_{j_T}&=&
\sum_{n=0}^{\infty} n \left\{{\cal F}_n(j_T\, |\, j_T-1,0; j_T+1,0)
\right. \nn \\
&&
\label{enct1}
\left. 
+{\cal F}_n(j_T-1\,|\, j_T,0; j_T+1,0)+
{\cal F}_n(j_T+1\,|\, j_T-1,0; j_T,0)\right\}.
\ea
The r.h.s. of eq. (\ref{enct1}) can again be 
expressed in terms of the generating functions

\ba
{\langle n \rangle}_{j_T}&=&\frac{\partial}{\partial z} 
\left\{ {\cal F}(j_T\,|\, j_T-1,0; j_T+1,0;z)\right.
\nn \\
&&
\label{genenct}
\left.+{\cal F}(j_T-1\,|\, j_T,0; j_T+1,0;z)
+{\cal F}(j_T+1\,|\, j_T-1,0; j_T,0;z) \right\}
\Big|_{z=1},
\ea
where 

\begin{equation}
{\cal F}(s\,|\,s_1,n_1;s_2,n_2;z)=\sum_{n=0}^{\infty} 
{\cal F}_n(s\,|\, s_1,n_1; s_2, n_2)\,z^n.
\end{equation}
To compute this generating function, let us first observe that 

\begin{equation}
\label{relPF}
{\cal P}_n(s\,|\, s_1,0; s_2,0)=\sum_{k=1}^n {\cal F}_k(s\,|\,s_1,0;s_2,0)\,
{\cal P}_{n-k}(0\,|\,s_1-s,0;s_2-s,0), \qquad s\neq 0,s_1,s_2,
\end{equation}
where ${\cal P}_n(s\,|\, s_1,0; s_2,0)$ is the probability that the 
walker is at site $s$ at time $n$ conditioned to 
its not having visited sites $s_1$ and 
$s_2$ but regardless of any previous visits to $s$. 
This probability must fulfil the initial condition 
${\cal P}_0(0\,|\,s_1-s,0;s_2-s,0)=1$. 
A simple calculation shows that eq. (\ref{relPF}) leads to the relation 

\begin{equation}
\label{genFgenP}
{\cal F}(s\,|\,s_1,0;s_2,0;z)=\frac{{\cal P}(s\,|\,s_1,0;s_2,0;z)}{
{\cal P}(0\,|\,s_1-s,0;s_2-s,0;z)},
\end{equation}
where 

\begin{equation}
{\cal P}(s\,|\,s_1,n_1;s_2,n_2;z)=\sum_{n=0}^{\infty} 
{\cal P}_n(s\,|\, s_1,n_1; s_2, n_2)\,z^n.
\end{equation}
This generating function can be computed in terms 
of ${\cal P}(s,z)$. For details we refer to the book by Weiss 
(ref. \cite{weiss}, chapter 4). The result for $n_1=n_2=0$ is 

\begin{equation}
\label{genP}
{\cal P}(s\,|\,s_1,0;s_2,0;z)={\cal P}(s,z)-
\sum_{k=1}^2 \frac{ {\cal D}_k(0)}
{{\cal D}(0)}\,{\cal P}(s-s_k,z),\quad s\neq s_1,s_2,
\end{equation}
where ${\cal D}_k(0)$ and ${\cal D}(0)$ are the following determinants:

\ba
&& {\cal D}(0)=\left| \begin{array}{cc}
{\cal P}(0,z) & {\cal P}(s_2-s_1,z) \\
{\cal P}(s_1-s_2,z) & {\cal P}(0,z) 
\end{array}
\right|,  \\
&& {\cal D}_1(0)=\left| \begin{array}{cc}
{\cal P}(s_1,z) & {\cal P}(s_2-s_1,z) \\
{\cal P}(s_2,z) & {\cal P}(0,z) 
\end{array}
\right|, \\
&& {\cal D}_2(0)=\left| \begin{array}{cc}
{\cal P}(0,z) & {\cal P}(s_1,z) \\
{\cal P}(s_1-s_2,z) & {\cal P}(s_2,z) 
\end{array}
\right|. 
\ea
We can now use these expressions to compute explicitly the mean collision
time ${\langle n \rangle}_{j_T}$. From eqs. (\ref{genenct}) and 
(\ref{genFgenP}) we have

\ba
{\langle n \rangle}_{j_T}&=&\frac{\partial}{\partial z} 
\left\{ \frac{{\cal P}(j_T\,|\,j_T-1,0;j_T+1,0;z)}
{{\cal P}(0\,|1,0;N-1,0;z)}
+\frac{{\cal P}(j_T-1 \,|\,j_T,0;j_T+1,0;z)}{
{\cal P}(0\,|\,1,0;2,0;z)}\right. \nn \\
&&
\label{enctgenfu}
\left. +\frac{{\cal P}(j_T+1\,|\,j_T-1,0;j_T,0;z)}{
{\cal P}(0\,|\,N-1,0;N-2,0;z)} \right\}\Bigg|_{z=1}
\ea
(recall that the arguments of the different generating functions 
are evaluated modulo $N$). Thus, the reaction time can again
be expressed in terms of ${\cal P}(j,z)$. This quantity is easily 
computed from the relation  



\begin{equation}
\label{singstprob}
{\cal P}_{n+1}(j)=\sum_{j'=0}^{N-1}{\cal P}_n(j')\hat{p}(j-j'),
\end{equation}
where the $\hat{p}(j)$'s are the single step probabilities for the 
random walk. 
Next, let us introduce the Fourier transform of the single step 
probabilities: 

\begin{equation} 
\tilde{p}\left(\frac{2\pi l}{N}\right)= 
\sum_{k=0}^{N-1}\hat{p}(k)\exp{\left(\frac{2\pi ik\,l}{N}\right)}.
\end{equation}
This expression can be used to represent ${\cal P}(j,z)$ as 
follows \cite{weiss}:

\begin{equation}
\label{Fougen}
{\cal P}(j,z)=\frac{1}{N}\sum_{k=0}^{N-1}\frac{\exp{(2\pi ikj/N)}}{
1-z\tilde{p}(2\pi k/N)},
\end{equation}
where $\tilde{p}(\cdot)$ is computed from the following 
single step probabilities:


\begin{equation}
\label{singstprob2}
\hat{p}(2)=\hat{p}(-2)=\frac{p}{4},
\quad \hat{p}(1)=\hat{p}(-1)=\frac{1-p}{2},\quad \hat{p}(0)=\frac{p}{2}.
\end{equation} 

%
We now have the necessary ingredients to evaluate the 
functions ${\cal P}(j_1\,|\,j_2,0;j_3,0;z)$ explicitly and to compute 
${\langle n \rangle}_{j_T}$ and $\langle n\rangle$ via the relations 
(\ref{genP}) and (\ref{enctgenfu}). 
The initial-condition-averaged reaction time $\langle 
n\rangle$ is obtained by averaging over a uniform 
ensemble of all possible non-reactive configurations characterized
by distinct values of $j_T$, i.e. 

\begin{equation} 
\label{avenct}
\langle n \rangle \equiv \frac{1}{N-3}
\sum_{j_T=2}^{N-2}{\langle n \rangle}_{j_T}.
\end{equation}


\subsection{Higher order moments}
\label{himo}

An important advantage of our generating function approach is 
that once ${\cal F}(i\,|\, j,0; k,0;z)$ is
known, the computation of higher order moments can be carried
out straightforwardly by deriving with respect to $z$. 
Indeed, one has

\ba
{\langle n^m \rangle}_{j_T}&=&\left( z\, \frac{\partial}{\partial z}\right)^m 
\left\{ {\cal F}(j_T\,|\, j_T-1,0; j_T+1,0;z)\right.
\nn \\
&&
\label{genmom}
\left.+{\cal F}(j_T-1\,|\, j_T,0; j_T+1,0;z)
+{\cal F}(j_T+1\,|\, j_T-1,0; j_T,0;z) \right\}
\Big|_{z=1},
\ea

In particular, the second-order moment  
 
\ba
{\langle n^2 \rangle}_{j_T}&\equiv&
\sum_{n=0}^{\infty} n^2 \left\{{\cal F}_n(j_T\, |\, j_T-1,0; j_T+1,0)
+{\cal F}_n(j_T-1\,|\, j_T,0; j_T+1,0)\right. \nn \\
&&
\left. +
{\cal F}_n(j_T+1\,|\, j_T-1,0; j_T,0)\right\}
=\left( z\, \frac{\partial}{\partial z}\right)^2
\left\{ {\cal F}(j_T\,|\, j_T-1,0; j_T+1,0;z)\right.
\nn \\
&&
\label{genvar}
\left.+{\cal F}(j_T-1\,|\, j_T,0; j_T+1,0;z)
+{\cal F}(j_T+1\,|\, j_T-1,0; j_T,0;z) \right\}
\Big|_{z=1},
\ea
the variance 

\begin{equation}
\langle v\rangle_{j_T}\equiv {\langle n^2 \rangle}_{j_T}-
{\langle n \rangle}_{j_T}^2,
\end{equation}
and its average over the different initial conditions 

\begin{equation}
\langle v\rangle\equiv \frac{1}{N-3}\sum_{j_T=2}^{N-2}\langle v\rangle_{j_T}
\end{equation}
are easily obtained. 
\section{Results}
\label{discuss}

\subsection{Behavior of the reaction time} 

\begin{figure}[ht]
\includegraphics[width=8cm,height=8cm]{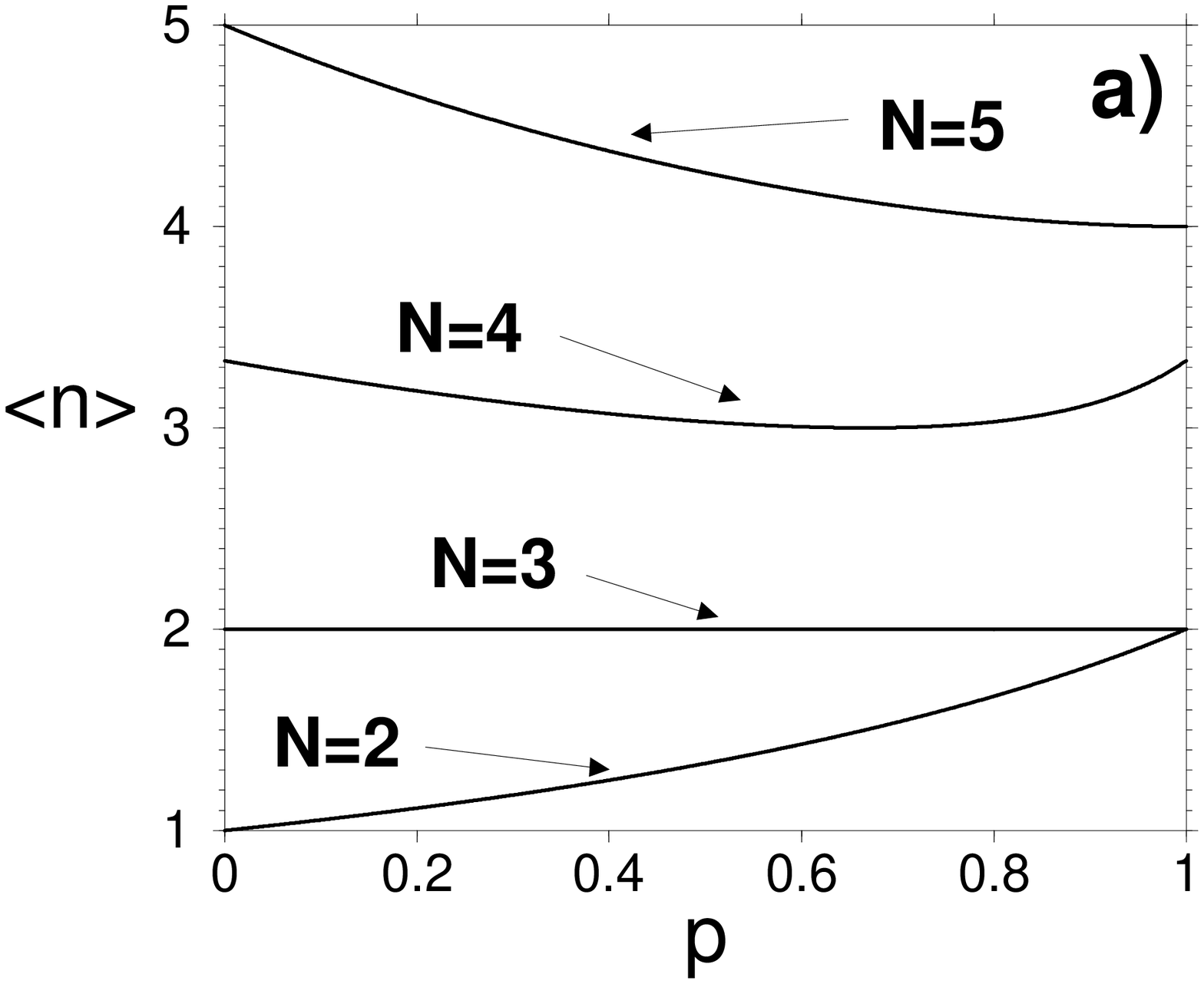}
\includegraphics[width=8cm,height=8cm]{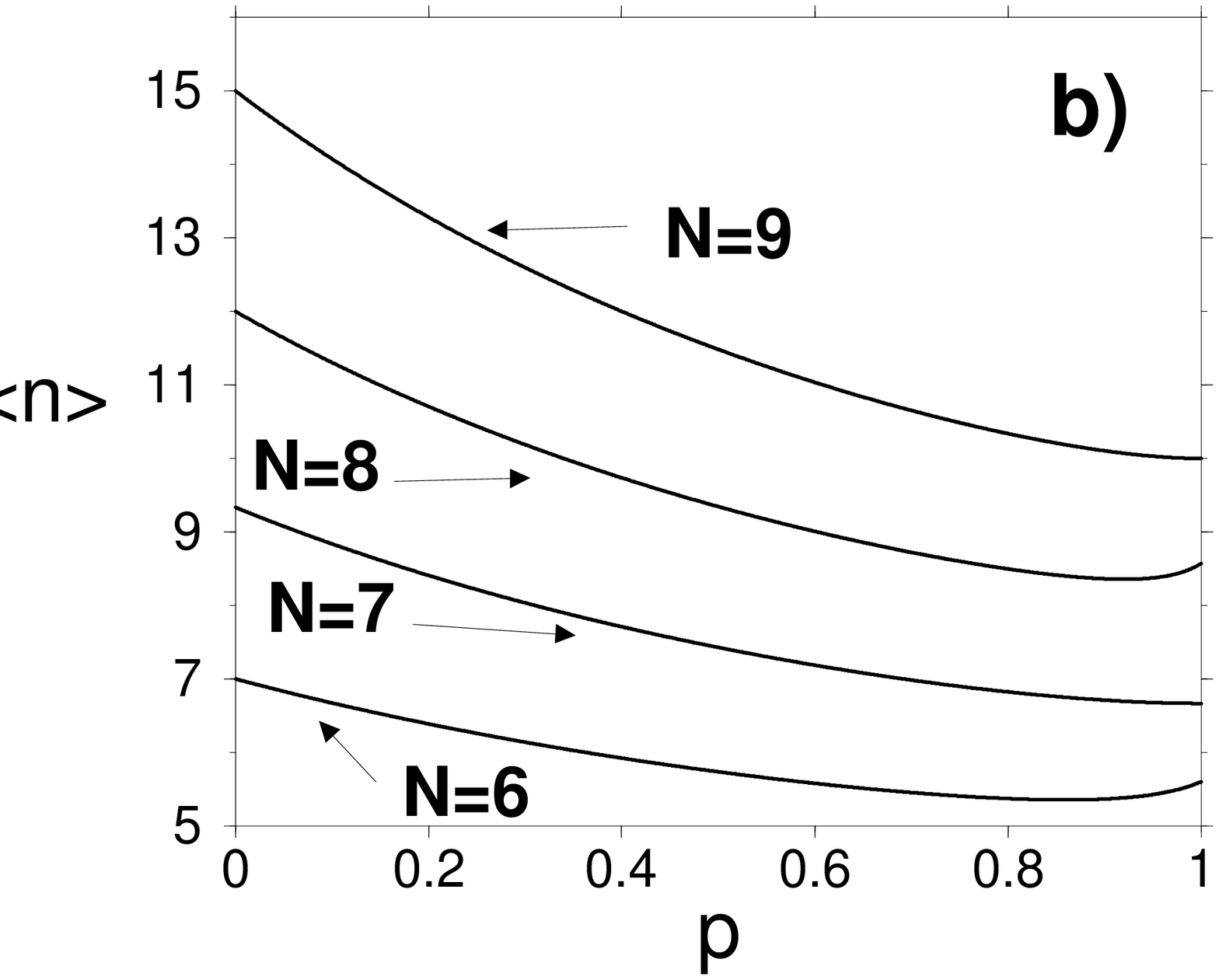}
\caption{\it \label{felfig8} Mean reaction time $\langle n\rangle$
as a function of 
$p$ for a) $N=2,\ldots,5$ and b) $N=6,\ldots,9$.}
\end{figure}

In order to obtain the results for model I, the lattice size
$N$ must now be decreased by two units in the expressions 
for $\langle n\rangle$ and $\langle v\rangle$ 
obtained in the framework of the above
formalism. 
The expressions for $\langle n\rangle$ as a function of $p$ for small
values of $N$ \footnote{For large values of $N$,
the evaluation of the generating function becomes rather time-consuming,
except in the limiting cases $p=0$ and $p=1$.}
are displayed in Table \ref{tabenct}. 
It is seen that these are ratios of two polynomials whose complexity 
grows with increasing lattice size. 
These results are in full agreement with previous findings for 
$\langle n\rangle$ obtained in the framework of a difference equation 
approach by the author and coworkers \cite{abad}. 
Fig. \ref{felfig8} displays the $\langle n \rangle$- plots  
computed from the ratios of polynomials given in Table \ref{tabenct}
for small values of the lattice size $N$. Let us now summarize their
main properties and provide an intuitive explanation for the observed 
behavior. 
 

The plots confirm
the validity of the expressions (\ref{purasc}) and (\ref{evenoddenct})
for the two limiting cases $p=0$ and $p=1$. 
For a fixed value of $p$ the collision time always increases with the
lattice size. On the other hand, the behavior of $\langle n\rangle$ as 
a function of $p$ is not always monotonic. For $N=2$ 
(the smallest physically interesting lattice) 
the most effective process is the purely asynchronous one,
and $\langle n\rangle$ increases monotonically with $p$.
For $N=3$ the reaction time does not depend on $p$. 
For $N=4,6,8$ the reaction 
time decreases with increasing $p$ in a wide regime of $p$ values, 
but it then increases again for a sufficiently large value of
$p$. In contrast, for $N=5,7$ and $9$ $\langle n\rangle$ decreases
monotonically with $p$. 

These results as well as numerical simulations for larger lattices 
confirm the existence of an even-odd effect for $N>3$. For odd 
values of $N$, $\langle n\rangle$ decreases monotonically with $p$, 
whereas for even values of $N$ it decreases up to a value 
$p_{min}$ and it increases monotonically beyond this value. 
The value $p_{min}$ is rapidly shifted to one as $N$ increases, and 
the curves get closer and closer to the linear law $\langle n\rangle
=\alpha (1+p)$ (with $\alpha>0$) predicted by the continuum 
approximation \cite{abad}. Deviations from this behavior for small
lattices, where the purely synchronous process is not always the 
most efficient one, may thus be regarded as an indication of the 
important role played by finite-size effects as well as by the 
spatial discretization imposed by the lattice.
 
The even-odd effect described above is reminiscent of the one observed
in the limiting case $p=1$ \cite{bentz,nickoz1}.
In the case $p=0$, only 
one of the reaction channels (SSO) will be open regardless of the
value of $N$. In contrast, when $p=1$ one has two distinct
behaviors depending on the parity of $N$. For odd values of 
$N$, both channels are open. However, for even values of $N$ 
reactions occur via a single channel 
for a given initial condition, 
i.e. through SSO when the distance $d$ is even or by NNC otherwise. 
Now, when $p$ is no longer strictly equal to one but still close to this
value, both reaction channels should be open, 
but the above parity effect still holds in a statistical 
sense, since NNC (SSO) reactions will be rare for an even (odd)
value of $d$. Hence the parity-dependent behavior of 
the curves $\langle n \rangle (p)$ in this regime.

The non-monotonic behavior of $\langle n\rangle$ for even values of $N$
can be explained in terms of a competition between synchronous and 
asynchronous events. While synchronous transport is the most efficient
mechanism to bring distant walkers in the vicinity of each other, it does not
always maximize the efficiency of the reaction once the walkers 
are within the typical interaction radius. In a prereactive
configuration with the walkers sitting at next to nearest neighbor sites,
reaction within the next step can only take place via a synchronous event.
However, if they are at nearest neighbor sites, reaction through an 
asynchronous event becomes possible and it is then
more efficient than reaction
through a synchronous event (the reaction takes place 
2 times out of 4 vs. 1 time out of 4).   
Thus, if $d$ and $N$ are such that 
prereactive configurations with contiguous particles are favored, 
the resulting competition between synchronous and asynchronous dynamics
will lead to an antiresonance of the reaction time (maximal efficiency)
for a value $p_{min}$ between 0 and 1. The statistical weight of such 
configurations is apparently stronger in the case of an even lattice, 
thus the antiresonance effect still prevails after averaging over the 
initial conditions, leading
to the observed non-monotonic behavior of $\langle n\rangle$. 
In the limit of large $N$, the role of diffusional transport becomes 
increasingly important for the efficiency, thus one has 
$p_{min}\to 1$. On the other hand, 
one can also show that the parity 
effect is washed out in the diffusive limit, where the SSO and
the NNC reaction channels become indistinguishable \cite{abad}.  

\subsection{Behavior of the variance}

\begin{figure}[ht]
\includegraphics[width=8cm,height=8cm]{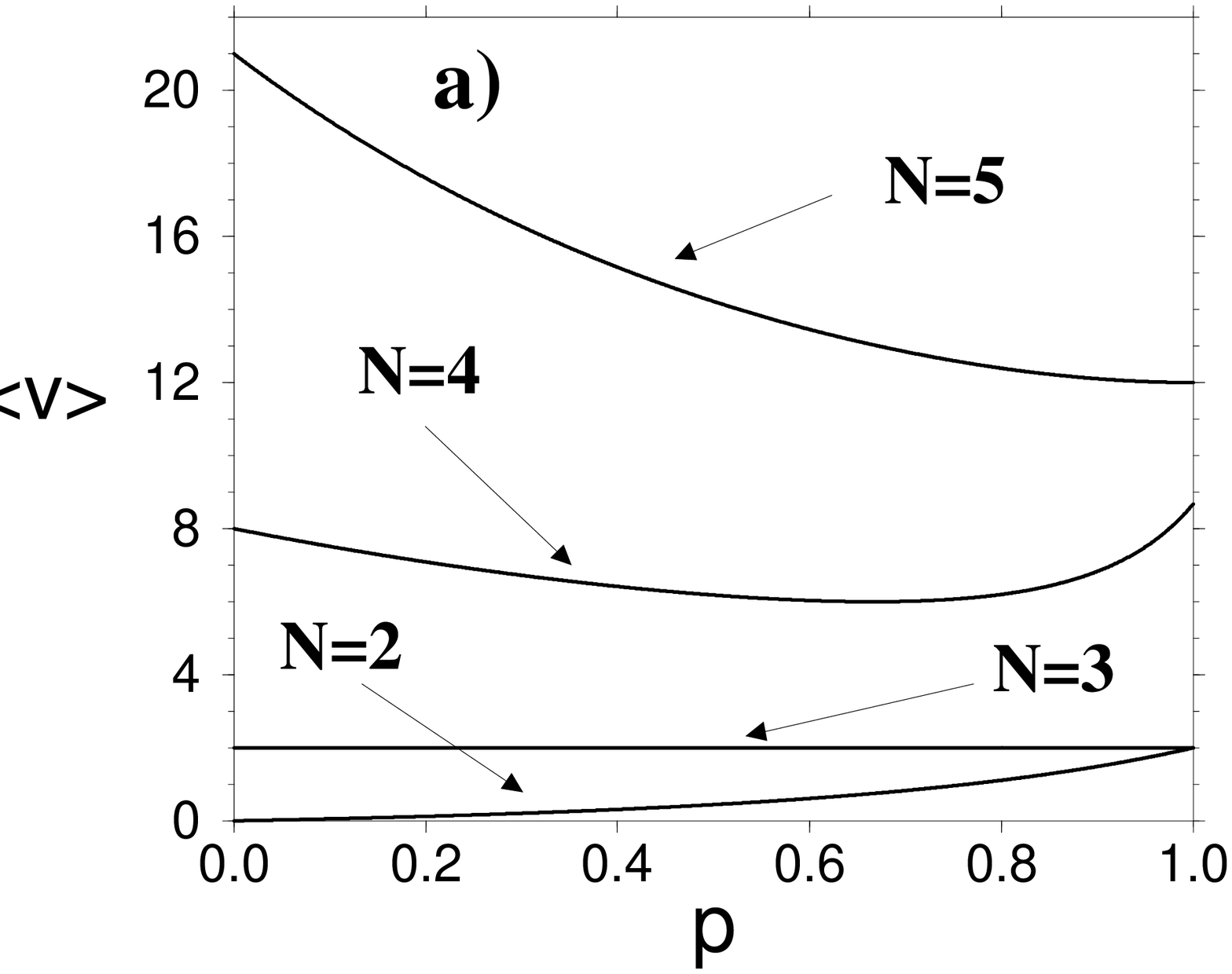}
\includegraphics[width=8cm,height=8cm]{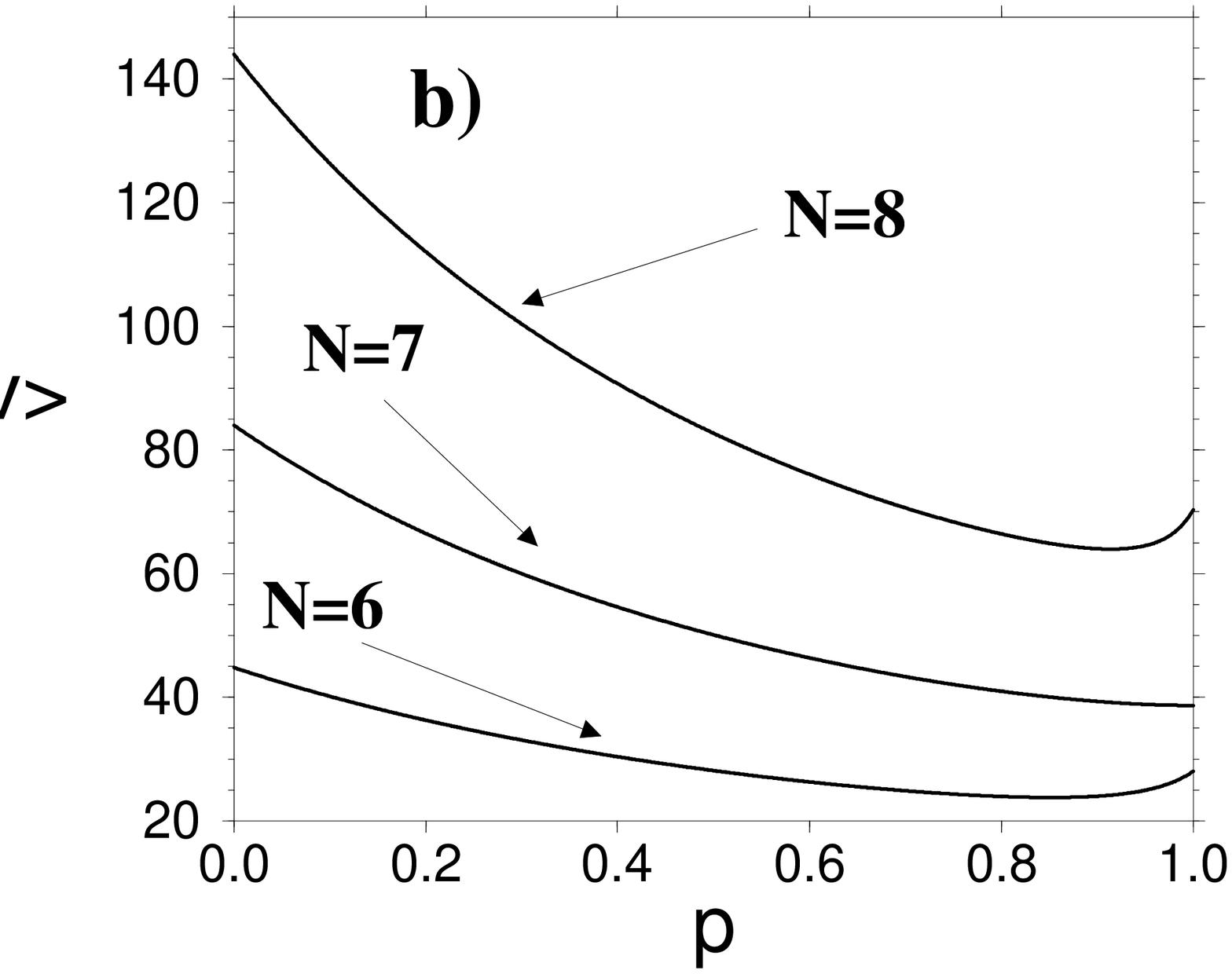}
\caption{\it \label{felfig9} Spatially averaged variance 
$\langle v\rangle$ of the encounter time 
as a function of $p$ for a) $N=2,\ldots,5$ and b) $N=6,\ldots,8$.}
\end{figure}

The analytic expressions for $\langle v\rangle$ as a function of $N$ in
model I are given in table \ref{tabvar}. 
The $p$-behavior of these functions follows essentially the same law as
$\langle n \rangle$ (cf. fig \ref{felfig9}). For odd $N$ the behavior is 
monotonically decreasing, while for even $N$ a minimum is observed 
which rapidly shiftes to $p=1$ with increasing $N$. This suggests that 
the behavior observed
for $\langle n\rangle$ extends to higher order moments; note, however, 
that the value $p_{min}$ is slightly
smaller than the one obtained from the $\langle n \rangle$-curves, except
for $N=4$, where it turns out to be the same ($p_{min}=2/3$). Thus, 
$\langle v\rangle$ and $\langle n\rangle$ cannot be simultaneously 
minimized except in this case.   

In the limiting cases $p=0$ and $p=1$ it is possible to simplify 
the expressions of the relevant generating function 
and thereby obtain general expressions for $\langle v\rangle$ 
for arbitrary lattice sizes. In the purely 
asynchronous case one has

\begin{equation}
\label{onewvar}
\langle v\rangle=\frac{N(N+1)(N-2)(N+2)}{30}.
\end{equation}
This expression is also recovered by taking the average of  
Montroll's original result for the variance over 
all possible initial walker-trap separations $d$ \cite{mont1}.  
On the other hand, the case $p=1$ yields

\begin{equation}
\label{evenoddvar}
\langle v\rangle=
\left\{ \begin{array}{cc}
N(N+1)(N+2)(N^2+2N+2)/(120 (N-1)) & \qquad \mbox{for $N$ even, } \\
(N+1)(N+3)(N^2+2N-5)/120 & \qquad \mbox{for $N$ odd. } 
\end{array} \right. 
\end{equation}

The standard deviation $\sigma\equiv\sqrt{\langle v\rangle}$ thus turns out 
to be comparable to $\langle n\rangle$ in both cases. According 
to eqs. (\ref{onewvar}) and (\ref{evenoddvar}), one has respectively
$\sigma\propto N^2/\sqrt{30}$ and $\sigma\propto N^2/(2\sqrt{30})$
for large $N$, while eqs. (\ref{purasc}) and (\ref{evenoddenct}) give
respectively $\langle n\rangle \propto N^2/6$ and 
$\langle n\rangle \propto N^2/12$. The standard deviation and the mean
value remain comparable for intermediate values of $p$. This is not surprising
in view of the large variability characteristic of first-passage problems.

\section{Conclusions}
\label{concl}

We have used a generating function approach to compute the mean
reaction time between two walkers performing a combination of 
synchronous and asynchronous jumps. The walkers react via two
channels, i.e. same site occupation or position exchange. The reaction 
time and its
variance display a different behavior for even and odd lattices, i.e.
they behave monotonically as a function of the synchronicity 
parameter $p$ for odd lattices and display antiresonances in the
even-lattice case. This behavior has been explained in terms of a
competition between synchronous and asynchronous dynamics. While the 
former favors diffusional transport over long distances, it
may also lead to a decrease of the cross section once the particles
are within the typical interaction radius, since they may avoid each other
more easily when they hop simultaneously. The even-odd effect tends to
vanish with increasing lattice size and in the continuum
limit, since it is a signature of the discrete nature of
the support. 

As pointed out in the Introduction,
the interest of considering the effect of synchronicity on the behavior
of a given system arises in a variety of different contexts. A fundamental
motivation is provided by the fact that, while free particles move
simultaneously under the action of a natural law at the microscopic 
level of description, in many real systems 
the existence of geometric and
energetic constraints (e.g. activation energies) 
results in an intrinsic nonzero degree of 
asynchrony in the dynamics of the constituents at mesoscopic 
time scales. Since in this context asynchronous dynamics can 
be understood as an expression of such constraints, it then
becomes natural to ask how it affects the efficiency of a 
given physical process. In view of our results, the answer
to this question cannot be considered to be straightforward. Moreover, 
it may turn out to be rather counterintuitive.  

Admittedly, the results depend strongly on
the specific definition of the collision rules. Yet, 
the equivalence between models I and II suggests that
certain features of the observed behavior might be characteristic
of a class of small systems. 
Similar antiresonance phenomena in
the encounter time have e.g. been recently observed in a model for 
target site localization of a protein on DNA \cite{copp}.  
On the other hand, the model shows that classical techniques
inspired in first-passage problems can be successfully used to 
compute characteristic reaction times for complex processes 
involving more than one interaction channel.  

Possible extensions of the model include the higher
dimensional case \footnote{In this case, there is numerical evidence for
an enhancement of the even-odd effect \cite{abad}.}, 
the case of more complex media
\cite{bujan,saxton,Bala}, and the evaluation of other quantities
such as survival probabilities \cite{anl,weiss4}, 
nearest-neighbor distance to the trap
\cite{weiss3} and concentration decays \cite{benpriv}
in the framework of the many-particle problem \cite{fish0,fish}
or more ellaborate reaction schemes \cite{walsh2,nickoz2}.




\section{Acknowledgements}

The author thanks Prof. G. Nicolis, 
J. Bentz, Prof. J. J. Kozak and Dr. V. Basios for helpful 
discussions.

\newpage


\begin{table}[ht]
\begin{tabular}{cc}
$N$ & $\langle n \rangle $ \\[0.5cm]
2  & $ 2/(2-p)$ \\
3 & 2 \\
4 & $(10/3)(3p-4)/(p^2+2p-4)$ \\
5 & $4 (2p-5)/( p^2-4)$ \\
6 & $(28/5) (p^2-10p+10)/(p^3-4p^2-4p+8) $ \\
7 & $(4/3) (p^2+8p-14)/(p^2-2)$ \\
8 & $(12/7) (13p^3+6p^2-126p+112)/((p-2)(p^3+6p^2-8))$ \\
9 & $10 (2p^3-5p^2-16p+24)/((p^2+2p-4)(p^2-2p-4))$ \\ 
10 & $(22/9) (7p^4-76p^3+16p^2+288p-240)/(p^5-6p^4-12p^3+32p^2
+16p-32)$ \\
\end{tabular}
\caption{\it \label{tabenct} Analytic expressions for 
$\langle n \rangle$ for different lattice sizes in model I.}
\end{table}

\begin{table}[ht]
\begin{tabular}{cc}
$N$ & $\langle v \rangle $ \\[0.5cm]
2  & $ 2p/(2-p)^2$ \\
3 & $2$ \\
4 & $(2/3)(192-316p+152p^2-15p^3)/(p^2+2p-4)^2$ \\
5 & $4(-2p^3+25p^2-80p+84)/(p^2-4)^2$ \\
6 & $(28/5)(-1096p+768p^2-206p^3+28p^4-p^5+512)/(8-4p+p^3-4p^2)^2 $ \\
7 & $(4/3)(-320p+88p^2+8p^3+p^4+252)/(p^2-2)^2$ \\
8 & $(12/7)(-13p^7+252p^6+546p^5-4736p^4-2184p^3+35520p^2-50848p+21504)
/((p^3+6p^2-8)^2(p-2)^2)$\\
9 & $2(-10p^7+253p^6-816p^5-2748p^4+11872p^3+5552p^2-42496p+29568)
/((p^2-2p-4)^2(p^2+2p-4)^2)$ \\ 
\end{tabular}
\caption{\it \label{tabvar} Analytic expressions for 
$\langle v \rangle$ for lattice sizes up to $N=9$ in model I.}
\end{table}

\begin{thebibliography}{99}

\bibitem{bouch} J.-P. Bouchaud and A. Georges, Phys. Rep. {\bf 195}, 127
(1990). 

\bibitem{benhav} D. ben-Avraham and 
S Havlin, {\it Diffusion and Reactions in Fractals and Disordered 
Systems}, Cambridge University Press, Cambridge (2000). 










\bibitem{abad} E. Abad, G. Nicolis, J. L. Bentz and 
J. J. Kozak, Physica A {\bf 326}, 69 (2003).

\bibitem{feller} W. F. Feller, {\it An Introduction to Probability Theory 
and its Applications}, Vol. I, Wiley, New York, 3rd edition (1968). 

\bibitem{kem} J. E. Kemeny and J. L. Snell, {\it Finite Markov chains}, 
Van Nostrand, Princeton (1960).

\bibitem{bentz} J. L. Bentz, J. J. Kozak, E. Abad and G. Nicolis, 
  Physica A {\bf 326}, 55 (2003).

\bibitem{these} E. Abad, PhD thesis, {\it Aspects of Nonlinear Dynamics
in low dimensional lattices: a multilevel approach}, 
Universit\'e Libre de Bruxelles (2003).

\bibitem{weiss} 
G. H. Weiss, {\it  Aspects and applications of the random walk},
North-Holland Elsevier Science, Amsterdam (1994).

\bibitem{redn} 
S. Redner, {\it A Guide to First-Passage Processes}, 
Cambridge University Press, Cambridge (2001).  


\bibitem{Ovch} A. A. Ovchinnikov and Ya. B. Zel'dovich, Chem. Phys. 
{\bf 28}, 215 (1978). 

\bibitem{tous} D. Toussaint and F. Wilczek, J. Chem. Phys. {\bf 78}, 
2642 (1983).

\bibitem{soko} I. M. Sokolov, H. Schn\"orer and A. Blumen, Phys. Rev. A
{\bf 44}, 2388 (1991).

\bibitem{lind} K. Lindenberg, B. J. West and R. Kopelman, Phys. Rev. A
{\bf 42}, 890 (1990). 

\bibitem{kang} K. Kang and S. Redner,  Phys. Rev. Lett.
{\bf 52}, 955 (1984).

\bibitem{kang2} K. Kang and S. Redner,  Phys. Rev. A
{\bf 32}, 435 (1985). 

\bibitem{mont0} E. W. Montroll,
Proc. Symp. Appl. Math. Am. Math. Soc. {\bf 16}, 193 
(1964).

\bibitem{mont1} E. W. Montroll, J. Math. Phys. {\bf 10}, 753 (1969).

\bibitem{monwei} E. W. Montroll and G. H. Weiss, J. Math. Phys.
{\bf 6}, 167 (1965).  

\bibitem{nickoz1} J. J. Kozak, C. Nicolis and G. Nicolis, J. Chem. Phys. 
{\bf 113}, 8168 (2000). 





\bibitem{copp} M. Coppey, O. Benichou, R. Voituriez and M. Moreau,
Biophys. J. {\bf 87}, 1640 (2004). 

\bibitem{bujan} M.C. Buj\'an Nu$\tilde{\mbox{n}}$ez, A. Miguel Fern\'andez
and M. A. Lopez Quintela, J. Chem. Phys. {\bf 112}, 8495 (2000). 

\bibitem{saxton} M. J. Saxton, J. Chem. Phys. {\bf 116}, 203 (2002). 

\bibitem{Bala} J. J. Kozak and V. Balakrishnan, Phys. Rev. E
{\bf 65}, 021105 (2002).

\bibitem{anl} J. K. Anlauf, Phys. Rev. Lett. {\bf 52}, 1845 (1984). 

\bibitem{weiss4} G. H. Weiss, S. Havlin and A. Bunde, 
J. Stat. Phys. {\bf 40}, 191 (1985). 

\bibitem{weiss3} G. H. Weiss, R. Kopelman and S. Havlin, Phys. Rev. A
{\bf 39}, R466 (1989). 

\bibitem{benpriv} D. ben Avraham, V. Privman and D. Zhong, Phys. Rev. E {\bf 
52}, 6889 (1995). 

\bibitem{fish0} M. E. Fisher, J. Stat. Phys. {\bf 34}, 667 (1984).

\bibitem{fish} M. E. Fisher and M. P. Gelfand, J. Stat. Phys. 
{\bf 53}, 175 (1988). 

\bibitem{walsh2} C. A. Walsh and J. J. Kozak, Phys. Rev. B
{\bf 26}, 4166 (1982).

\bibitem{nickoz2} C. Nicolis, J. J. Kozak and G. Nicolis, J. Chem. Phys. 
{\bf 115}, 663 (2001).

\end{thebibliography}
\end{document}